\documentclass[a4paper,fleqn]{cas-dc}

\usepackage[authoryear]{natbib}
\AtBeginDocument{\setlength{\FullWidth}{\textwidth}}

\def\tsc#1{\csdef{#1}{\textsc{\lowercase{#1}}\xspace}}
\tsc{WGM}
\tsc{QE}
\tsc{EP}
\tsc{PMS}
\tsc{BEC}
\tsc{DE}

\begin{document}
\let\WriteBookmarks\relax
\def\floatpagepagefraction{1}
\def\textpagefraction{.001}

\shorttitle{Towards a Likelihood Ratio Approach for Bloodstain Pattern Analysis}

\shortauthors{Tong Zou, Hal Stern}

\title [mode = title]{Towards a Likelihood Ratio Approach for Bloodstain Pattern Analysis}                      
\tnotemark[1]

\tnotetext[1]{This work was funded (or partially funded) by the Center for Statistics and Applications in Forensic Evidence (CSAFE) through Cooperative Agreements 70NANB15H176 and 70NANB20H019 between NIST and Iowa State University, which includes activities carried out at Carnegie Mellon University, Duke University, University of California Irvine, University of Virginia, West Virginia University, University of Pennsylvania, Swarthmore College and University of Nebraska, Lincoln.}


%
\author[1]{Tong Zou}[orcid=0000-0003-1302-8363]
\cormark[1]


\ead{zout3@uci.edu}




\author[1]{Hal Stern}

\cortext[cor1]{Corresponding author}

\ead{sternh@uci.edu}

\address[1]{Department of {Statistics,University of California, Irvine},Irvine,CA,92617}




\begin{abstract}
In this work, we explore the application of likelihood ratio as a forensic evidence assessment tool to evaluate the causal mechanism of a bloodstain pattern. It is assumed that there are two competing hypotheses regarding the cause of a bloodstain pattern. The bloodstain patterns are represented as a collection of ellipses with each ellipses characterized by its location, size and orientation. Quantitative measures and features are derived to summarize key aspects of the patterns. A bivariate Gaussian model is chosen to estimate the distribution of features under a given hypothesis and thus approximate the likelihood of a pattern. Published data with 59 impact patterns and 55 gunshot patterns is used to train and evaluate the model. Results demonstrate the feasibility of the likelihood ratio approach for bloodstain pattern analysis. The results also hint at some of the challenges that need to be addressed for future use of the likelihood ratio approach for bloodstain pattern analysis.
\end{abstract}




\maketitle

\section{Introduction}
\label{intro}

Bloodstain pattern analysis (BPA) has been employed as a means of crime scene reconstruction and testimonial evidence for more than one hundred years \citep{james2005principles}. Frequently, bloodstain patterns encountered at a crime scene are evaluated to determine the mechanism behind the bloodletting event \citep{bevelrm}. By analyzing the shapes, sizes, orientations and locations of bloodstains along with other information, BPA experts develop hypotheses about on how the event happened. Another task of BPA is to determine the spatial location of the bloodletting event by inferring the trajectories of blood droplets \citep{camana2013determining, attinger2019determining}, but this is not addressed by the work reported here. Although BPA can play a critical role as a forensic tool, its accuracy and reliability have been questioned during the last decade. A comprehensive study of forensic science by the National Research Council of the National Academy of Sciences expressed concerns. The report notes that ``\textit{the opinions of bloodstain pattern analysts are more subjective than scientific}'' and ``\textit{The uncertainties associated with bloodstain pattern analysis are enormous}'' \citep{national2009strengthening}. Current BPA approaches rely heavily on techniques taught during workshops, and on experience and experimentation. The recently published black box study \citep{hicklin2021accuracy} finds several examples of bloodstain patterns for which BPA analysts do not agree on the mechanism, and others where the majority of analysts conclude an incorrect mechanism. 

There has been recent interest in developing quantitative approaches to BPA. The hope is that such approaches may allow for more scientific analyses that appropriately address uncertainty. The most common approach to date has attempted to build models to classify bloodstain patterns as being due to one of a few possible causal mechanisms \citep{arthur2017image, arthur2018automated, liu2020automatic}. In these studies, a number of quantitative features are devised based on an ellipse representation of bloodstain patterns. Then a classifier is trained using those features to distinguish between two causal mechanisms, e.g. impact vs cast-off \citep{arthur2018automated} and gunshot vs impact \citep{liu2020automatic}. A significant limitation of the classification framework is that a pattern will always be classified into one of the trained categories, even if it is actually caused by a different mechanism. It is not generally appropriate to say with certainty that a pattern is produced by one of only two mechanisms.

An alternative to the classification approach is the likelihood ratio (LR) approach. The LR is a concept in statistics designed as a measure of degree to which observed data is more likely under one hypothesis as against an alternative hypothesis. In the context of forensics, the LR can be used to quantitatively estimate the relative support provided by the evidence for two competing propositions of forensic interest \citep{stern2017statistical}. A number of authors have advocated for the LR approach \citep{aitken2004statistics, willis2015strengthening}. Attempts have been made to apply the LR approach to many kinds of forensic evidence like DNA \citep{steele2014statistical}, latent prints \citep{neumann2012quantifying}, digital \citep{galbraith2017analyzing, galbraith2020quantifying}, firearms \citep{bunch2013application} and handwriting \citep{bozza2008probabilistic, marquis2011handwriting, gaborini2017towards}. We briefly explain the approach here. Suppose evidence is to be evaluated under two competing hypotheses $H_1$ and $H_2$. By definition LR is the ratio of the probabilities of the evidence conditioning on $H_1$ and $H_2$, in formula:
\begin{align}
LR = \frac{P(E|H_1)}{P(E|H_2)} \label{lrdef}
\end{align}

A value of the LR greater than one implies that the evidence is more likely under $H_1$ than under $H_2$ and therefore provides more support for $H_1$ than $H_2$, and vice versa for a value of the LR smaller than one. Larger (or smaller) value of the LR provides stronger support for $H_1$ (or $H_2$). Various authors have provided interpretations for different values of the LR (see, e.g., \citealp{kass1995bayes}). It is important to note that the value of the LR greater than one does not indicate that $H_1$ is more probable that $H_2$. Such a statement would be referring to posterior probabilities $P(H_1|E)$ and $P(H_2|E)$, which require specifying prior probabilities. For more on this point see \citet{aitken2004statistics}. The likelihood evaluation process for many evidence types models the similarity between a crime scene item and a reference item. The corresponding hypotheses concern whether they are from the same source or not. This is where bloodstain pattern analysis diverges from other forms of forensic evidence. The BPA setting does not involve assessing between the evidence and a reference item. Recently, \citet{attinger2022using} reviewed the potential for applying the LR approach in BPA. They note the complexities, describe needed development and demonstrate via a hypothetical example regarding the time of an event. This article provides a proof of concept for the LR approach in the context of mechanism determination.

The hypotheses $H_1$ and $H_2$ are assumed to represent hypotheses a bloodstain pattern is produced under two different mechanisms. For example, in the David Camm case \citep{hicklin2021accuracy}, $H_1$ and $H_2$ can be taken to represent the hypotheses that bloodstain pattern is caused by gunshot or by transfer. The example we focus on, based on data availability, is discrimination of gunshot patterns and impact patterns. As noted in \citep{liu2020automatic}, this remains a challenging and meaningful problem in forensics. Based on this scenario, we show how the LR framework can be used to analyze bloodstain evidence.

In order to obtain the LR, we need to be able to estimate $P(E|H_1)$ and $P(E|H_2)$ in a quantitative way given a bloodstain pattern $E$. These are certainly cases where one can be confident that a pattern is produced by one of two mechanisms, but this needs not generally be the case. In the Appendix \ref{Appendix} we show that even when one considers a wide range of potential mechanisms, it is still relevant to model $P(E|H)$ for each of the potential mechanisms. In terms of modeling the likelihood of a bloodstain pattern, our approach consists of three main steps: first, the bloodstain pattern image is processed and represented by a collection of ellipses approximating the constituent stains. We applied the technique from the work of \citet{zou2021recognition} to infer the number of ellipses and their numerical characteristics of the ellipses. Second, a few predefined quantitative features are extracted from the ellipse representation of each pattern. The features are all dimensionless quantities such that they are highly robust and can be easily adapted for other data sets. Third, the distribution of the features is estimated for a set of bloodstain patterns produced by the same mechanism and used as the likelihood model. By demonstrating the potential use of the LR framework for bloodstain patterns, we aim to provide insights for future researchers and legal professionals as they work to incorporate statistical tools into reasoning about bloodstain evidence.

Section \ref{Data Preparation} describes the data we use, characteristics of the images and the image processing methods. Section \ref{method} describes the statistical approach to evaluating the LR in detail. This includes the definition of features and the development of probability models that estimate the likelihood under different mechanisms. Application of the approach to data from impact and gunshot patterns \citep{attinger2018data, attinger2019data} is described in Section \ref{Experiments}. Discussion including limitations and challenges is presented in Section \ref{Discussion}.

\section{Data Preparation}
\label{Data Preparation}

\subsection{Bloodstain Images}
\label{Bloodstain Images}

Two collections of bloodstain images are published as open source data sets, with one containing impact patterns \citep{attinger2018data} and the other gunshot spatters \citep{attinger2019data}. An impact pattern is created when an object strikes a blood source. A gunshot pattern is created when the blood source is shot by a bullet and the stains generated from high-velocity droplet projectiles. In this study we use the 59 single-source impact patterns and the 55 gunshot patterns on vertical targets from these data sets.

Here we briefly describe the experiments designed to generate the patterns. All patterns are produced indoors using fresh swine blood at room temperature. A vertical target wall covered with cardstock or butcher paper collects the flying blood droplets to form a pattern. Since the scanner window used to create the images is smaller than the target area, patterns are scanned piece-wise at a resolution of 600dpi and then stitched together using a graphics editor. 

\begin{figure*}[ht]
	\centering
	  \includegraphics[width=\textwidth]{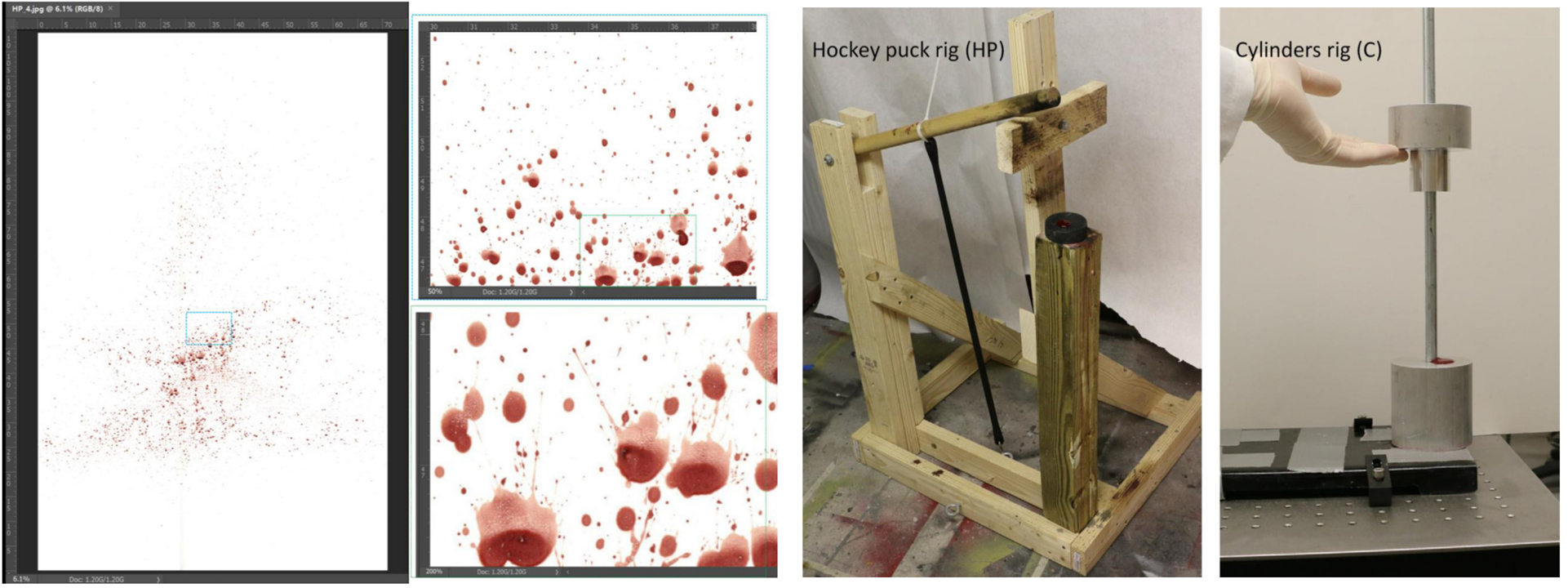}
	\caption{Images taken from \citep{attinger2018data}. On the left is an impact pattern on a $70\times 110$ cm board  with different scales of insets. On the right are the two apparatuses used to create impact patterns.}
	\label{impact}
\end{figure*}

Two apparatuses are used to create the impact patterns. In the first, a blood pool around 1 mL is hit by a metal cylinder weight released from different heights above the blood. This simulates stepping into puddles of blood. In the second apparatus, the blood pool is placed on a hockey puck and hit by a thin wooden rod with different velocities, imitating beating incidents with a blunt weapon. The horizontal distance between the blood source and target wall is varied between 30 and 200 cm. Figure \ref{impact} shows a sample impact pattern and pictures of the two apparatuses.

\begin{figure*}[ht]
	\centering
	  \includegraphics[width=\textwidth]{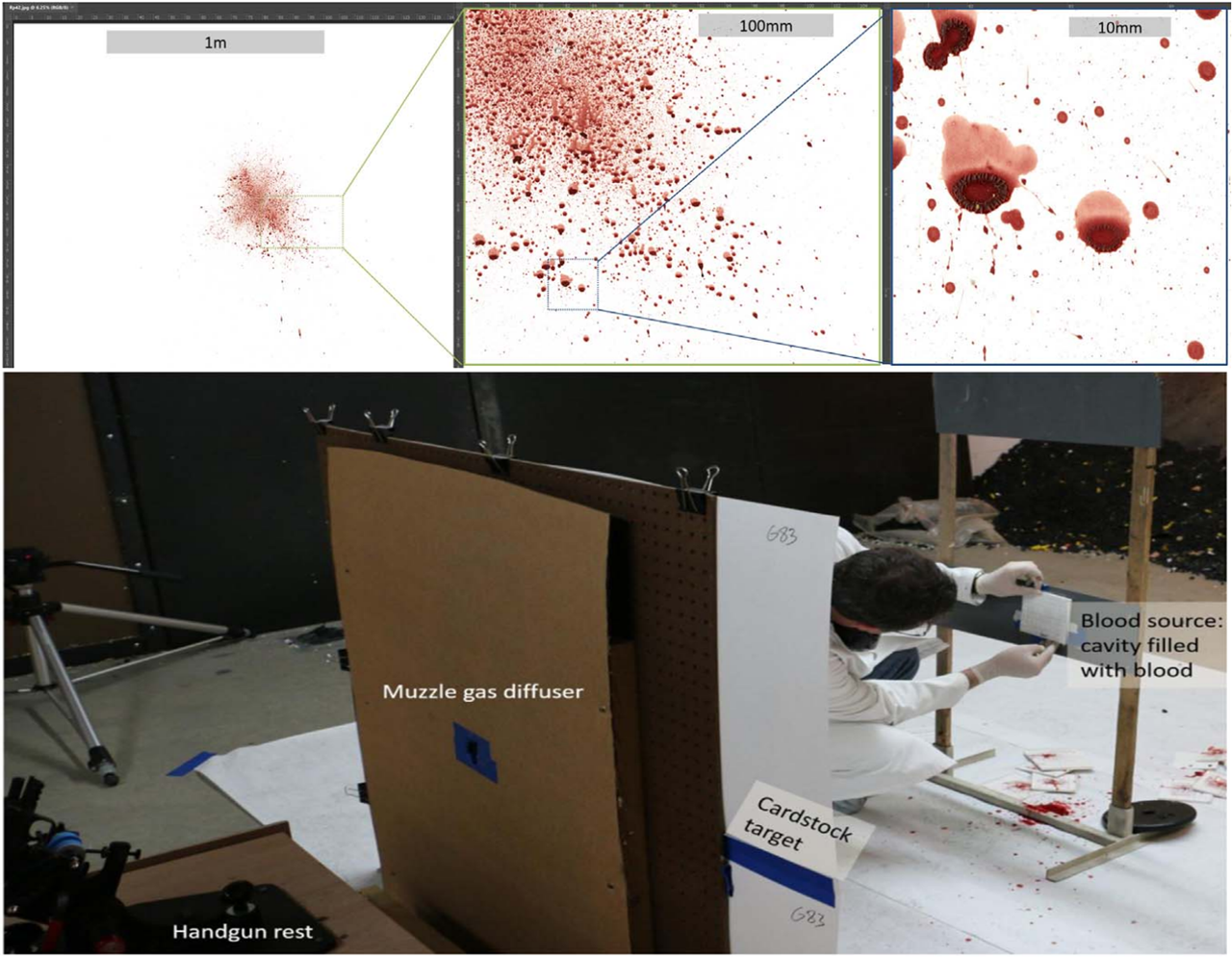}
	\caption{Images taken from \citep{attinger2019data}. On the top is a gunshot pattern on a $140\times 110$ cm cardstock board with different scales of insets. On the bottom is the experiment preparation of a backspatter with a handgun.}
	\label{gunshot}
\end{figure*}

The gunshot patterns are all back spatters where bloodstains are splashed in the opposite direction to that of the bullet. The source is a piece of foam filled with blood. It is then shot by a bullet traveling horizontally, creating a pattern on the target wall between the source and the gun. To add variability, handguns and rifles with different bullet shapes (pointy, round, and flat) are used. The horizontal distance between the target wall and blood source is varied between 10 and 120 cm. Figure \ref{gunshot} shows a sample gunshot pattern and a picture of the laboratory setting.

\subsection{Image Processing}
\label{Image Processing}

We assume the evidence $E$ in formula (\ref{lrdef}) in a bloodstain pattern analysis is an image of the crime scene. Directly modeling the distribution of a bloodstain pattern with a limited number of samples is beyond the bounds of possibility due to the high-dimensionality of image data. For impact and gunshot patterns, many of the bloodstains in a pattern are in the shape of ellipses. Considering a pattern as a collection of ellipses can greatly simplify the data structure and retain most of the information in the pattern. Thus, the first step in our analysis is processing the images to obtain an ellipse representation for each bloodstain pattern.

The JPEG images are imported in MATLAB R2021 \citep{MATLAB:2021b} and pre-processed using the DIPImage Toolbox v2.9 \citep{dipimage}. We follow the procedures described in the work of \citet{arthur2017image} to transform the original RGB image to a smoothed binary image. The transformation is carried out through a series of steps: \textit{background subtraction}, \textit{element segmentation} and \textit{morphological operations}. First, the image is converted to grayscale and an inferred background is subtracted out. The background image is estimated through downsampling of the original image, median filtering, and upsampling to the original size. Then, the image is converted to black and white (the segmentation step) using the Triangle thresholding algorithm \citep{zack1977automatic}. Lastly, binary erosion and dilation each with four iterations are applied to eliminate image noise and remove any tail pixels associated with individual bloodstains.

\begin{figure*}[ht]
	\centering
	  \includegraphics[width=\textwidth]{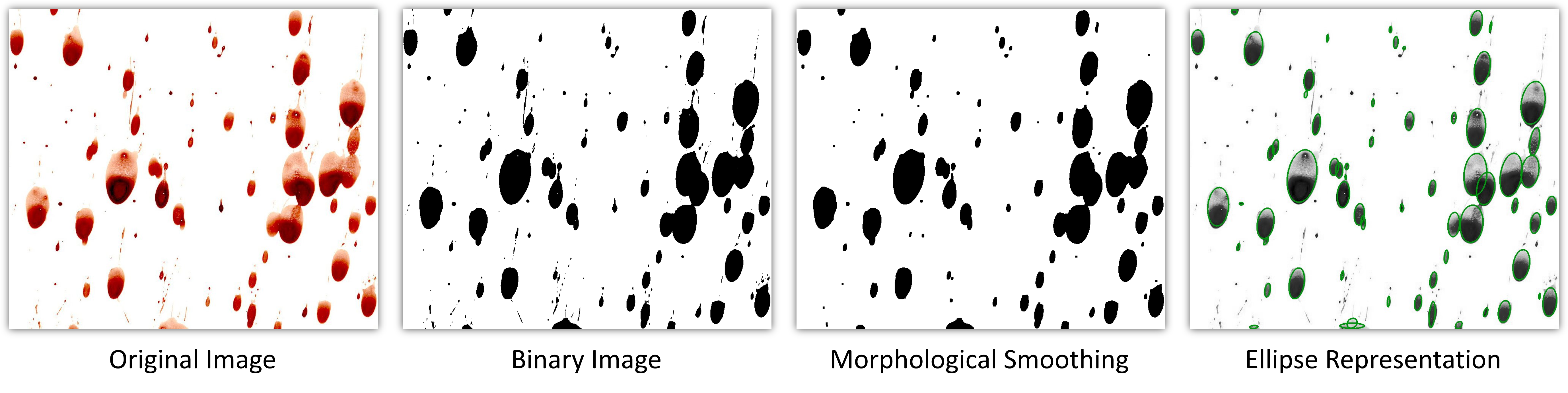}
	\caption{An outline of image processing. The binary image is obtained after thresholding. The morphological smoothed image is obtained after binary erosion and dilation. The ellipse representation is produced by the ellipse recognition algorithm.}
	\label{outline}
\end{figure*}

After those steps, Arthur et al. use the \textit{regionprops} function (in MATLAB's Image Processing Toolbox) to label and measure every region of the binary image. In particular, each region is fitted by an ellipse to approximate the shape of a blood droplet, and parameters of the ellipse are later used to define features. A limitation of this method is that \textit{regionprops} only fits each region with one ellipse while regions that may be composed of multiple droplets are discarded, which inevitably results in a loss of information. Instead, we use the algorithm developed in \citet{zou2021recognition} that can approximate a non-elliptical region with a set of overlapping ellipses. The idea of the algorithm is first to generate a pool of candidate ellipses through a series of distance transforms of the region, then partition the region contour into several segments defined by its concave points, and finally match each segment with an ellipse from the candidate pool. Regions that are purely noise or poorly approximated by ellipses are excluded using the following approach. Two metrics, the Jaccard Index \citep{jaccard1912distribution} and Hausdorff distance \citep{huttenlocher1993comparing}, are computed to assess the fit between each region and its approximating ellipses (or a single ellipse). A region is discarded if either of the metric exceeds a specified threshold (0.9 for Jaccard Index and 5 for Hausdorff distance). An example of image processing is provided in Figure \ref{outline}. As a final step, any bloodstain pattern that is represented by five ellipses or fewer is discarded as not having sufficient data. The processing results in 59 impact patterns and 52 gunshot patterns (i.e., 3 gunshot pattern images are removed). 

\begin{figure}[ht]
\centering
\resizebox{0.3\textwidth}{!}{%
\includegraphics{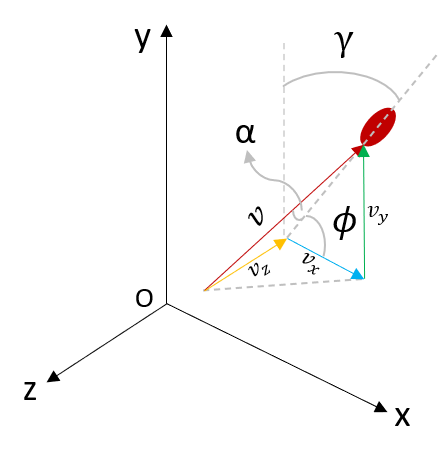}
}
\caption{Coordinate system and orientation angles associated with a blood droplet.}
\label{notation}
\end{figure}

Now that bloodstains are approximated by a collection of ellipses, we use the parameters of these ellipses as a quantitative representation of a pattern. An ellipse can be characterized by five parameters $(x,y,a,b,\phi)$, where $(x,y)$ denote the center coordinate, $a$ and $b$ the radii of major and minor axes, and $\phi$ the angle between the x-axis and the major axis of the ellipse (see Figure \ref{notation}). A pattern is represented by an $n$ by 5 table where $n$ is the number of approximating ellipses and each row contains the ellipse parameters.

\section{Method}
\label{method}

Given the available data introduced in Section \ref{Bloodstain Images}, we specify the two competing hypotheses regarding a bloodstain pattern to match the two mechanisms of the datasets. $H_{1}$ and $H_{2}$ in formula (\ref{lrdef}) now propose that the bloodstain pattern is produced by gunshot and impact respectively as follows:
\begin{align*}
H_1&:\textit{The pattern is produced by gunshot} \\ H_2&:\textit{The pattern is produced by impact}
\end{align*}
In Section \ref{Image Processing} we transform a bloodstain pattern image into a collection of ellipses, or equivalently, an $n$ by 5 table filled with values of ellipse parameters. Despite the great reduction in data complexity, the data table is still in a high-dimensional space and modeling the likelihood is not straightforward with a limited number of samples. We further simplify the problem by modeling the likelihood of a number of features $\boldsymbol{f}$ that are summaries of the ellipse representation. As a result, the likelihood ratio of a given pattern is defined by the following:
\begin{align}
LR_{BPA} = \frac{P(\boldsymbol{f}|H_1)}{P(\boldsymbol{f}|H_2)}
\end{align}

It is worth mentioning that our goal in this study is to illustrate how the likelihood ratio approach can be used to assess the relative support for competing hypotheses about the mechanism that produced a given bloodstain pattern. The limited data available and simplified model indicate that this is a demonstration of the approach rather than a definitive LR calculation.

\subsection{Feature Design}
\label{FD}

Previous studies \citep{arthur2018automated, liu2020automatic} have proposed a number of quantitative summary features useful for classifying patterns from different mechanisms. Most of the features rely on the ellipse representation. Examples include the average area of ellipses, the average ratio of the lengths of the major and minor axes and the number of ellipses. Both those studies identify the orientation of bloodstains as an important characteristic of a pattern and therefore proposed descriptive statistics sensitive to orientation as features. For instance, bloodstains in cast-off patterns tend to have more consistent orientation than those in impact patterns. This suggests that the variance of the slopes of the ellipses in an impact pattern is likely to be larger than that of a cast-off pattern. However, as discussed in the following section, because of the inherent property of periodicity of orientation variables, it is not appropriate to use the traditional definition of variance in statistics. Instead, we applied directional statistics to solve the problem of periodicity and designed new features that exploit the expected importance of orientation of bloodstains.

\subsubsection{Circular Features}
\label{CF}

Two commonly used measures to gauge the orientation of an individual bloodstain are the gamma angle and the impact angle. As shown in Figure \ref{notation}, the gamma angle, $\gamma$, is the angle between the blood droplet path projected on the target surface ($Z=0$) and the plumb (vertical) line. The impact angle $\alpha$ is the angle between the blood droplet path and the target surface, With $\gamma$ and $\alpha$, one can deduce the direction of velocity of a blood droplet when it hits the surface. In terms of the ellipse representation, one can estimate the impact angle using the formula:
\begin{align}
\quad \alpha = \arcsin{\frac{b}{a}}
\end{align}
As for the gamma angle, it seems intuitive to express $\gamma$ via ellipse slope $\phi$ since they are complementary angles. However, due to the fact that an ellipse is symmetric in both of its major and minor axes, $\phi$ cannot determine the head and tail of a blood droplet as it ranges from 0 to 180 degrees whereas $\gamma$ can range from 0 to 360 degrees. \citet{arthur2017image} proposed comparing the centroids of a bloodstain before and after the morphological operations to determine its directionality (head and tail). This method is applicable only to single-droplet stains with a connected tail, while the directionality of stains without a noticeable tail or overlapped with multiple droplets are difficult to infer. In this study, we choose to use ellipse slope $\phi$ in place of the gamma angle in order to include more bloodstains into the analysis.

Summary statistics like the variance of the distribution of $\phi$ and $\alpha$ have previously been used to characterize the orientation of ellipses \citep{arthur2018automated, liu2020automatic}. However, traditional statistics of angular variables actually fail to properly describe the population. For example, the average of a 2 degree angle and a 358 degree angle is $(2+358)/2=180$ degrees, which is far from the more sensible answer 0 degree. In principle, statistics should be invariant and consistent to coordinates change. But the average value of angles depends on the reference line that defines 0 degree and whether the angle ranges from 0 to 360 degrees or -180 to 180 degrees. This is because conventional statistics assumes all variables belong to the Euclidean space where a straight line never forms into a loop, but angular variables lie in a circular space. The subject that studies the statistical behavior of such variables is called directional statistics \citep{mardia2009directional}. A common approach is to map an angular variable $\beta$ ranging from 0 to 360 degrees into 2-dimensional space via polar coordinates:
\begin{align}
\beta \in [0, 360) \mapsto \boldsymbol{u} = (\cos{\beta}, \sin{\beta}) \in 	\mathbb{R}^2
\end{align}
Now the angular variables are distributed on a 2-dimensional unit circle. To obtain the average of a group of angles, we can instead average the corresponding vectors and compute the angle of the centroid vector as the angle average:
\begin{align}
\Bar{\boldsymbol{u}} &= \frac{1}{n}\sum_{i=1}^n \boldsymbol{u}_i = (\frac{1}{n}\sum_{i=1}^n \cos{\beta_i}\ , \ \frac{1}{n}\sum_{i=1}^n \sin{\beta_i}) \\
\Bar{\beta} &= \text{angle between }\Bar{\boldsymbol{u}}\text{ and vector } (1, 0)
\end{align}
The length of $\Bar{\boldsymbol{u}}$ is called the \textit{mean resultant length} and denoted as $\Bar{R}$. The variance of $\beta$ is then defined as one minus the mean resultant length:
\begin{align}
Var[\beta] &= 1 - \lVert\Bar{\boldsymbol{u}}\rVert \label{eq_var} \\
&= 1 - \Bar{R} \\
&= 1 - \sqrt{(\frac{1}{n}\sum_{i=1}^n \cos{\beta_i})^2 + (\frac{1}{n}\sum_{i=1}^n \sin{\beta_i})^2} \label{cirvar}
\end{align}
It is obvious that $\Bar{R}$ is always less than one, so the variance of an angular variable ranges from zero to one. Considering the extreme cases makes it easier to understand why the variance is defined in this manner. If all $\beta_i$'s have the same value, which means there is no variation, $\Bar{R}$ will be exactly one and $Var[\beta]$ will be zero. On the other hand, if $\beta_i$'s are uniformly distributed between 0 and 360, which means maximum variation, $\Bar{\boldsymbol{u}}$ will be averaged into zero vector, leading $Var[\beta]$ to its maximum one. $\Bar{R}$ and $Var[\beta]$ reflects the degree of concentration and dispersion of angles respectively.

The variance of the impact angle $\alpha$ can be computed using formula (\ref{cirvar}). In terms of the slope angle $\phi$ for which 0 degree is the same as 180 degrees, it is a standard procedure to first multiply $\phi$ by a factor of two so that it ranges from 0 to 360 degrees to conform with the expected range for argument of trigonometric functions. Formulas used to calculating the variances of the two angles are given in Table \ref{feature}.

\subsubsection{Spherical Features}

Because the direction of incident velocity of a blood droplet can be deduced given $\alpha$ and $\gamma$ (or its complement $\phi$), features defined in terms of those variables (e.g. $Var[\phi]$ and $Var[\alpha]$) can summarize the distribution of incident direction of blood droplets to some extent. A limitation of this approach is that information about the correlation of these two angles is not incorporated.

Here we describe an approach based on viewing the data through vectors of incident direction. This will allow us to incorporate info about the correlation. Note that we can derive the vector of incident direction in three dimensions from $\gamma$ and $\alpha$:
\begin{align}
\boldsymbol{m} = \frac{\boldsymbol{v}}{\lVert\boldsymbol{v}\rVert} = (-\cos{\alpha}\cos{\gamma}, -\cos{\alpha}\sin{\gamma}, -\sin{\alpha})
\end{align}
where $\boldsymbol{v}$ is the incident velocity (Figure \ref{notation}) and $\boldsymbol{m}$ is a unit vector in the same direction as $\boldsymbol{v}$. Since the gamma angle has been replaced by $\phi$ as described in the previous section, we use the following instead:
\begin{align}
\Tilde{\boldsymbol{m}} = (-\cos{\alpha}\cos{2\phi}, -\cos{\alpha}\sin{2\phi}, -\sin{\alpha})
\end{align}
where the factor of 2 before $\phi$ is to maintain its circular structure as explained above. The vector $\Tilde{\boldsymbol{m}}$ does not necessarily equal $\boldsymbol{m}$ due to the use of $\phi$. But the distribution of $\Tilde{\boldsymbol{m}}$ still precisely characterizes the variation of incident directions of blood droplets in the case that their head and tail directions are unknown.

Although $\Tilde{\boldsymbol{m}}$ is a 3-dimensional vector, by its construction from the two angles, it is constrained to lie on the surface of a hemisphere with unit radius centered at origin. Due to the non-Euclidean nature of the spherical surface, conventional statistical methods fail to function properly. The strategies we used in the previous section to handle angular variables is applicable to spherical variables. For example, the mean resultant length of $\Tilde{\boldsymbol{m}}$, defined as:
\begin{align}
\Tilde{R} = \lVert\frac{1}{n}\sum_{i=1}^n \Tilde{\boldsymbol{m}}_i\rVert
\end{align}
is still a good measure of concentration. But since $\Tilde{\boldsymbol{m}}$ is 3-dimensional, a more informative and accurate alternative is to compute the scatter matrix $\boldsymbol{T}$:
\begin{align}
\boldsymbol{T} = \sum_{i=1}^n \Tilde{\boldsymbol{m}}_i \Tilde{\boldsymbol{m}}_i^T \label{scatterT}
\end{align}
In mechanics, $\boldsymbol{T}$ is interpreted as the inertia tensor of a rigid body about the origin, where the body is composed of equal weight particles at each of the locations $\Tilde{\boldsymbol{m}}_1, ..., \Tilde{\boldsymbol{m}}_n$ with the same weight. In statistics, $\boldsymbol{T}$ is analogous to the covariance matrix that provides information regarding variances and correlations of data. One possible summary to capture the dispersion of $\boldsymbol{T}$ is its determinant $\det(\boldsymbol{T})$, which in fact relates to the data entropy assuming the distribution is multivariate Gaussian \citep{ahmed1989entropy}.

\begin{figure*}[ht]
\centering
\resizebox{1\textwidth}{!}{%
\includegraphics{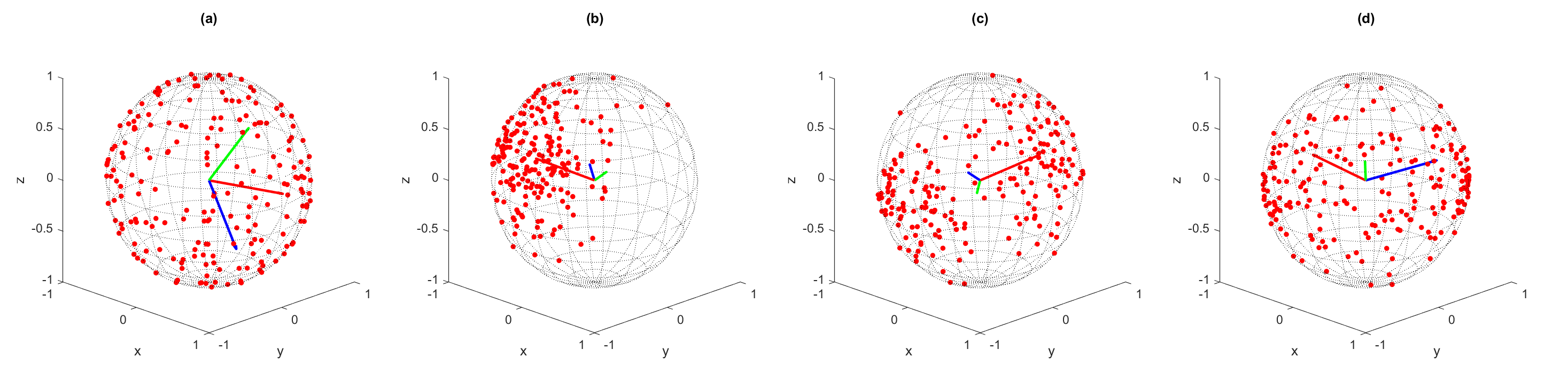}
}
\caption{Examples of spherical distributions described in Table \ref{tablesphere}. (a) Uniform. (b) Unimodal. (c) Bimodal. (d) Girdle. Data points (red dots) are scattered on the sphere surface. In each sphere there are three lines of red, blue and green indicating the directions of eigenvectors $\boldsymbol{t}^1$, $\boldsymbol{t}^2$, $\boldsymbol{t}^3$ respectively. Their lengths are proportional to $t^1$, $t^2$, $t^3$ and scaled so that the red line has the same length as radius.}
\label{sphdist}
\end{figure*}

Another way to more specifically characterize the shape of the distribution of incident directions is via an eigendecomposition of $\boldsymbol{T}$. The eigendecomposition of the covariance matrix provides a set of orthogonal principal components that are often used to extract features. Indeed, as pointed out in \citep{mardia2009directional}, the eigenvalues $t^1$, $t^2$, $t^3$ (in decreasing order) and eigenvectors $\boldsymbol{t}^1$, $\boldsymbol{t}^2$, $\boldsymbol{t}^3$ of $\boldsymbol{T}$ indicate the general shape of the data as described in Table \ref{tablesphere} and illustrated in Figure \ref{sphdist}. The eigenvalues of the scatter matrix are great candidates to serve as features for bloodstain patterns due to their mutual independence and summarization of the distribution of incident directions. The distribution of $\Tilde{\boldsymbol{m}}$ will most likely fall into the unimodal category in Table \ref{tablesphere}, because $\Tilde{\boldsymbol{m}}$ is constrained on only half of the sphere and categories with other shapes of distribution involve the whole sphere. Thus, $t^1$ is expected to be much greater than $t^2$ and $t^3$. The ratio between $t^2$ and $t^3$ indicates the symmetry of the distribution and can serve as an appropriate summary. 

\begin{table*}[ht]
\centering
\caption{Descriptive interpretation of shapes of spherical distribution based on the eigendecomposition of $\boldsymbol{T}$ \citep{mardia2009directional}}
\renewcommand{\arraystretch}{2}
\begin{tabular}{|c|c|c|}
\hline
Relative magnitudes of eigenvalues & Shape of distribution & Other features \\
\hline
$t^1 \approx t^2 \approx t^3$ & uniform & \\
\hline
$t^1$ large; $t^2$, $t^3$ small & & \multirow{3}{25em}{
if $t^2 \approx t^3$, rotational symmetry about $\boldsymbol{t}^1$ \\
otherwise, concentrated at $\boldsymbol{t}^1$ (and $-\boldsymbol{t}^1$ for bimodal)
} \\
(i) $\Bar{R} \approx 1$ & unimodal & \\ 
(ii) $\Bar{R} < 1$ & bimodal &  \\ 

\hline
\multirow{2}{9em}{$t^1$, $t^2$ large; $t^3$ small} & \multirow{2}{3em}{girdle} & \multirow{2}{25em}{if $t^2 \approx t^3$, rotational symmetry about $\boldsymbol{t}^3$   \\
otherwise, concentrated about circle in plane of $\boldsymbol{t}^1$, $\boldsymbol{t}^2$} \\ 
& & \\
\hline
\end{tabular}
\label{tablesphere}
\end{table*}

\subsection{Likelihood Evaluation}
\label{LE}

The LR framework involves computing the likelihoods of a bloodstain pattern under different assumptions about the mechanism. As described earlier, directly modeling the likelihood of a bloodstain pattern is extremely difficult. To demonstrate the approach, we instead model the joint likelihood of features extracted from the pattern. Theoretically, patterns that are caused by different mechanisms become more distinguishable if more features are used in the model. But this makes the result more difficult to interpret and harder to visualize. As a compromise, we model the joint distribution of two selected features $f_1$ and $f_2$ with a bivariate Gaussian distribution under $H_1$ and under $H_2$:
\begin{align}
\begin{pmatrix} f_1\\ f_2\\ \end{pmatrix}|H_1 \sim N_2(\boldsymbol\mu_1,\boldsymbol\Sigma_1), \  \begin{pmatrix} f_1\\ f_2\\ \end{pmatrix}|H_2 \sim N_2(\boldsymbol\mu_2,\boldsymbol\Sigma_2)
\end{align}
where the parameters $\boldsymbol\mu_1, \boldsymbol\Sigma_1$ are the mean and covariance of the distribution of features extracted from gunshot patterns, and $\boldsymbol\mu_2, \boldsymbol\Sigma_2$ are those parameters for features extracted from impact patterns. The choice of the Gaussian distribution may seem like a strong assumption, but works well for transformations of the features in our example as described in Section \ref{FP}. Mixture models and kernel density estimation are alternative approaches when the distribution of features is complex and hard to transform into a Gaussian distribution \citep{lindsay1995mixture,jones1996brief}. The LR of a bloodstain pattern is evaluated by first extracting its pair of features $\boldsymbol{f} = (f_1, f_2)^T$ and then computing the Gaussian density functions under the competing hypotheses:
\begin{align}
LR_{BPA} \approx \frac{p_{N_2}(\boldsymbol{f}|\boldsymbol\mu_1,\boldsymbol\Sigma_1)}{p_{N_2}(\boldsymbol{f}|\boldsymbol\mu_2,\boldsymbol\Sigma_2)} \label{lrbpa}
\end{align}
where $p_{N_2}(\boldsymbol{f}|\boldsymbol\mu,\boldsymbol\Sigma)$ is the density function of a bivariate Gaussian distribution. Parameters of the density functions are estimated by maximum likelihood on a training data set. For example, parameters of the model under $H_1$ are estimated as follows:
\begin{gather}
\hat{\boldsymbol\mu}_1 = \frac{1}{N}\sum_{j=1}^{N}\boldsymbol{f}_j \label{mu}\\
\hat{\boldsymbol\Sigma}_1 = \frac{1}{N}\sum_{j=1}^{N}(\boldsymbol{f}_j - \hat{\boldsymbol\mu}_p)(\boldsymbol{f}_j - \hat{\boldsymbol\mu}_p)^T \label{variance}
\end{gather}
where $\boldsymbol{f}_j$'s are features extracted from patterns that are produced by gunshot in our data set, and $N$ is the number of gunshot patterns. Parameters of the model under $H_2$ are estimated similarly using impact patterns. Note here we use pattern index $j$ and number of patterns $N$ to differentiate from the ellipse index $i$ and number of ellipses $n$ within a pattern. In practice, the model requires a much larger data set with high variations to accurately estimate the parameters. Despite the limitation of the simplified Gaussian model and the small data set, our work still shows promising results in the following section. Consequently, the simplicity helps to demonstrate the essence of the LR framework and the potential for implementation in BPA.

\section{Experiments}
\label{Experiments}

The objective of this section is to illustrate the LR approach presented in Section \ref{method} using the data introduced in Section \ref{Data Preparation}.

\subsection{Feature Preprocessing}
\label{FP}

Features derived in Section \ref{FD} are grouped into two pairs: (1) the circular features are $Var[\phi]$ and $Var[\alpha]$, the variances of two orientation angles; (2) the spherical features are the ratio between the third and second largest eigenvalues of the scatter matrix $\boldsymbol{T}$ (which measures symmetry) and the determinant of $\boldsymbol{T}$ (which measures dispersion). Descriptions and formulas to compute these features are give in Table \ref{feature}.

\begin{table*}[ht]
\centering
\caption{Characteristic features designed to represent bloodstain patterns}
\renewcommand{\arraystretch}{2}
\begin{tabular}{|c|c|l|c|c|}
\hline
Feature & Formula & Description & Range & Transform \\
\hline
\multirow{2}{4em}{\centering Circular \\ \centering Features} & $Var[\phi] = 1 - \sqrt{(\frac{1}{n}\sum_{i=1}^n \cos{2\phi_i})^2 + (\frac{1}{n}\sum_{i=1}^n \sin{2\phi_i})^2}$ & variance of gamma angle & $(0,1)$  & logit  \\
\cline{2-5}
 & $Var[\alpha] = 1 - \sqrt{(\frac{1}{n}\sum_{i=1}^n \cos{\alpha_i})^2 + (\frac{1}{n}\sum_{i=1}^n \sin{\alpha_i})^2}$ & variance of impact angle & $(0,1)$ & logit \\
\hline
\multirow{2}{4em}{\centering Spherical \\ \centering features}  & $t^3/t^2$  & symmetry of incident directions & $(0,1)$ & logit \\
\cline{2-5}
& $ \det(\boldsymbol{T}) = t^1 \times t^2 \times t^3$ & dispersion of incident directions & $(0,\infty)$ & log \\
\hline
\end{tabular}
\label{feature}
\end{table*}

\begin{figure*}[ht]
\centering
\resizebox{0.8\textwidth}{!}{%
\includegraphics{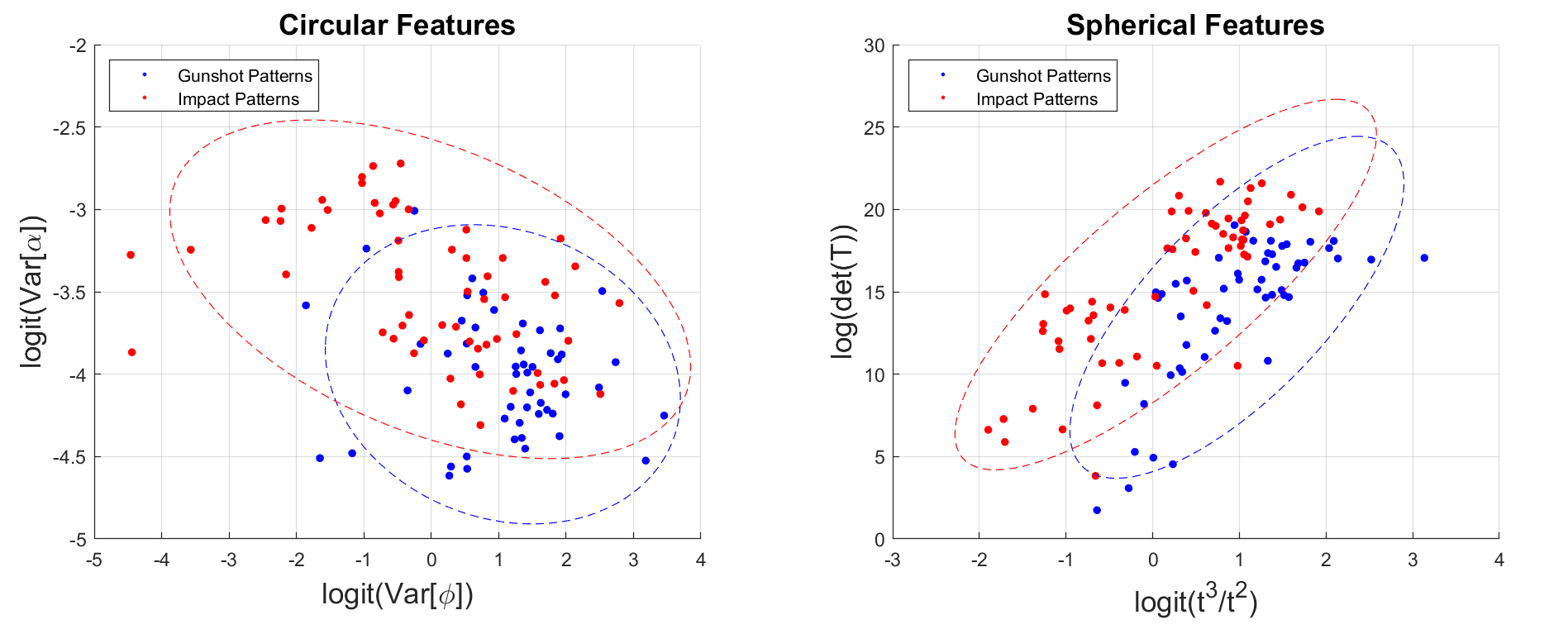}
}
\caption{Scatter plot of the transformed features. On the left are the circular pair of features for the patterns. On the right are the spherical pair of features for the patterns. A 95\% confidence ellipse is computed for each cluster.}
\label{scatter}
\end{figure*}

Our bivariate Gaussian model assumes the data distribution for the features is unbounded and not skewed. Therefore, instead of directly using these features in the model, we first map them into the real line $\mathbb{R}$ via transformations that should better approximate the assumed Gaussian distribution (see Table \ref{feature}). For example, $Var[\phi]$ ranging from 0 to 1 can be transformed to $\mathbb{R}$ by the logit function:
\begin{align}
\text{logit}(x) = \log(\frac{x}{1-x})
\end{align}
The transformed features extracted from all bloodstain patterns are plotted in Figure \ref{scatter}. The confidence ellipse defines the smallest region that contains 95\% of the points for each cluster. Due to transformations of the features, the shape of red and blue clusters in both plots roughly resemble bivariate Gaussian distribution. It is worth noting that the cluster of spherical features of gunshot patterns seems to consist of two sub clusters, which means a mixture model might fit the data better. But here we maintain the Gaussian model for simplicity. Gunshot patterns and impact patterns are not perfectly separated in either plot, though the degree of overlap between the two types of pattern (in red and blue colors) is obviously less for the spherical features than for the circular features.

On the left plot showing the circular features, we observe that gunshot patterns on average have smaller $Var[\alpha]$ and larger $Var[\phi]$. This indicates that in gunshot patterns, the impact angles of bloodstains are more converged and the slopes of ellipses are more dispersed than those in impact patterns. On the right plot showing the spherical features, we observe that incident directions are less disperse and more symmetric in gunshot patterns. Those observations may be explained by the fact that the average velocity of blood droplets produced by gunshot is higher than for those produced by impact.

\subsection{Model Evaluation}

To illustrate the LR calculation in a way that provides for a realistic evaluation of the results, we perform a leave-one-out cross validation. First, we select one bloodstain pattern from the data set as the test pattern. The rest of the patterns are used to estimate the parameters of the models using equations (\ref{mu}) and (\ref{variance}). Finally the selected test pattern is evaluated via the LR defined by formula (\ref{lrbpa}) with the estimated parameters to produce a LR value. This procedure is iterated until all patterns have been selected as the test pattern, thereby yielding $N = 111$ LR values. We expect gunshot patterns to produce LR values greater than 1, and impact patterns to have LR values less than one.

\begin{figure*}[ht]
\centering
\resizebox{\textwidth}{!}{%
\includegraphics{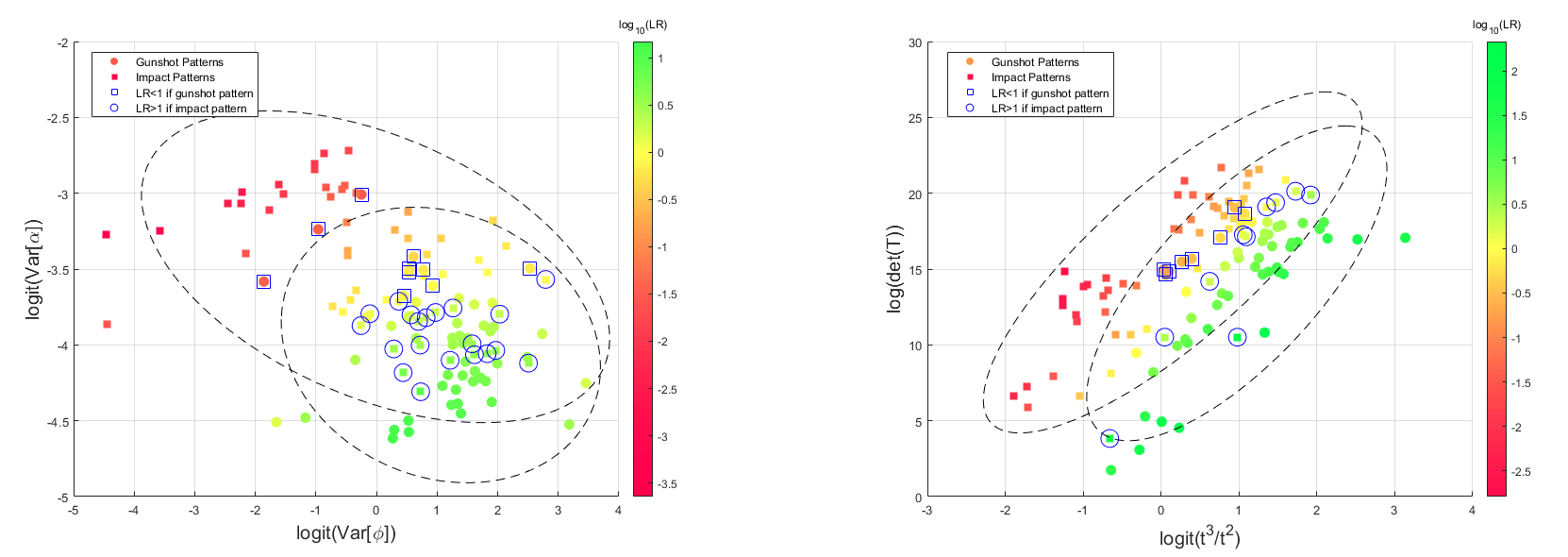}
}
\caption{Scatter plot of the circular features (left) and spherical features (right). The 95\% of confidence ellipses for impact and gunshot patterns are depicted in dash lines. Solid circles and squares represent gunshot and impact patterns respectively. The values of the log of the LR obtained from cross validation are expressed in color. Hollow circles and squares are superimposed on patterns where the LR supports the wrong hypothesis.}
\label{LRplot}
\end{figure*}

Figure \ref{LRplot} presents the LR values of patterns in color (actually the logarithm of the LR values) and highlights the cases for which the LR supports the wrong hypothesis. Points distant from the boundary between squares and circles have large absolute value of the log of LR. The LR of some patterns is greater than 100 or less than 0.01. Table \ref{confusion} reports the classification for all patterns using 1 as a threshold for assessing the LR. For both circular and spherical features, impact patterns being misclassified as gunshot patterns occurs more frequently than the reverse. Error rates for circular features and spherical features are respectively 27\% and 16\%. LR values near one indicate that the data are equally consistent with both hypotheses. Patterns with such values may contributed to these high error rates when we classify using a threshold of one. To examine whether the results are sensitive to the choice of the LR threshold, Table \ref{confusion1} gives the numbers of patterns with LR inside and outside the intermediate range from 0.5 to 2. There are more patterns assigned to the intermediate range for the model using circular features than for the model using spherical features. This indicates that the model of circular features yields LRs more concentrated around one and therefore more sensitive to the choice of LR threshold. As expected, the cases with $0.5<LR<2$ are evenly divided among the two mechanisms. These cases account for a number of errors in each analysis.

\begin{table}[ht]
\centering
\caption{Confusion matrices with $LR=1$ as threshold}
\renewcommand{\arraystretch}{1}
\begin{tabular}{|c|c|c|c|c|}
\hline
 \multirow{2}{4em}{\centering Ground \\ \centering Truth}  & \multicolumn{2}{c|}{Circular Features} & \multicolumn{2}{c|}{Spherical Features} \\
 \cline{2-5}
& $LR > 1$ & $LR < 1$ & $LR > 1$ & $LR < 1$ \\
  \hline
 $H_1$ & 42 & 10 & 44 & 8 \\
  \hline
 $H_2$ & 20 & 39 & 10 & 49 \\
   \hline
\end{tabular}
\label{confusion}
\end{table}

\begin{figure*}[ht]
\centering
\resizebox{0.85\textwidth}{!}{%
\includegraphics{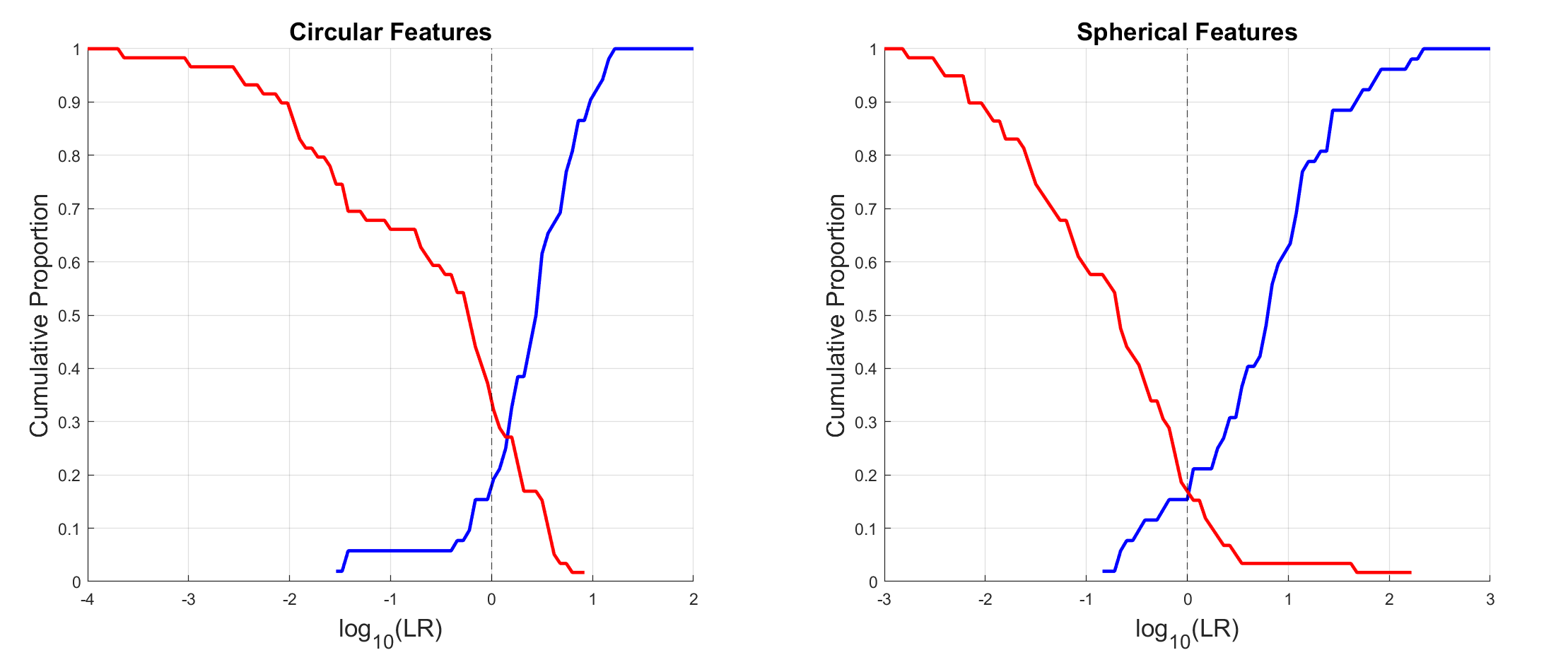}
}
\caption{Tippett plots for circular and spherical features. }
\label{tippett}
\end{figure*}

\begin{table*}[ht]
\centering
\caption{Confusion matrices with $0.5 < LR < 2$ as an intermediate zone}
\renewcommand{\arraystretch}{1}
\begin{tabular}{|c|c|c|c|c|c|c|}
\hline
 \multirow{2}{4em}{\centering Ground \\ \centering Truth}  & \multicolumn{3}{c|}{Circular Features} & \multicolumn{3}{c|}{Spherical Features} \\
 \cline{2-7}
& $LR > 2$ & $0.5 < LR < 2$ & $LR < 0.5$ & $LR > 2$ & $0.5 < LR < 2$ & $LR < 0.5$ \\
  \hline
 $H_1$ & 32 & 16 & 4 & 39 & 7 & 6 \\
  \hline
 $H_2$ & 10 & 22 & 27 & 5 & 7 & 39 \\
   \hline
\end{tabular}
\label{confusion1}
\end{table*}
\citet{morrison2021consensus} suggest the use of a Tippett plot to diagnose bias of the LR model. The Tippett plot portrays empirical cumulative probability distributions of LR under $H_1$ and $H_2$. In general, greater separation and gentler slopes of the two curves indicate better performance. We created the Tippett plot for both circular and spherical features, see Figure \ref{tippett}. The blue line gives the proportion of gunshot patterns (true $H_1$) with $log_{10}(LR)$ less than or equal to the corresponding value on the x axis. The red line gives the proportion of impact patterns (true $H_2$) with $log_{10}(LR)$ greater than or equal to the corresponding value on the x axis. The Tippett plot of circular features exhibits a slight shift to the right where the two curves intersect, meaning that impact patterns are more prone to misclassification than gunshot patterns with threshold $LR = 1$. The plot of spherical features has a flatter curve slope and almost no shift. In general, the spherical features demonstrate better performance than the circular features, indicating that the distribution of incident directions is more discriminating than the distribution of orientation angles.

\subsection{Misclassified Patterns}

\begin{figure*}[ht]
\centering
\resizebox{\textwidth}{!}{%
\includegraphics{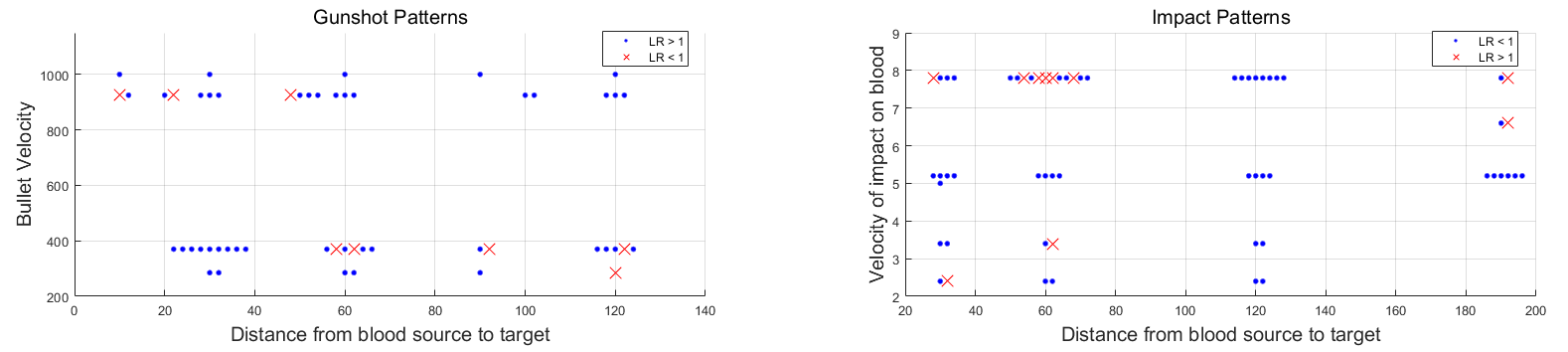}
}
\caption{Scatter plot of gunshot (left) and impact (right) patterns in terms of their conditions under which they were generated. The horizontal axis denotes the distance from the blood source to the target board. The vertical axis denotes different velocity levels of impact on blood when creating the pattern. Patterns where the LR obtained from the model of spherical features supports the wrong hypothesis are marked by red cross.}
\label{mis}
\end{figure*}

Analyzing patterns with LR values that are not consistent with their true mechanism can help us better understand and improve the model. There are two factors that were systematically controlled when the data were generated \citep{attinger2018data, attinger2019data}: (a) the distance from the blood source to the target board; (b) the initial velocity of the blood droplets. The spherical features show better performance than the circular features in the sense of having fewer misclassified patterns, so we choose to examine those patterns with respect to the two factors mentioned above. Figure \ref{mis} shows the factor level for each pattern and identifies the misclassified ones with a red cross according to LR values assigned by the model of spherical features. We observe that the majority of the misclassified impact patterns (8 out of 10) occurred at the higher velocity levels, while the majority of the misclassified gunshot patterns (5 out of 8) occurred at the lower velocity level. This result supports our conjecture in Section \ref{FP} that the observation of less dispersion of the impact angle and incident direction for the gunshot patterns is due to the greater velocity of blood droplets. Gunshot patterns with low velocity and impact patterns with high velocity may possess similar characteristic in terms of bloodstain orientation and hence are hard to distinguish for the model.

\section{Discussion}
\label{Discussion}

A likelihood ratio approach to evaluating bloodstain pattern evidence is presented as a demonstration of the potential for this approach. Building on the work of \citet{attinger2022using}, we address the problem of identifying the bloodletting mechanism. Based on two published data sets \citep{attinger2018data, attinger2019data}, statistical models of novel features are built to estimate the likelihood ratio of a bloodstain pattern to assess the relative support for the two possible mechanisms. Our work serves as a demonstration of the approach rather than a definitive likelihood ratio calculation as there are multiple ways in which it can be improved.

One limitation of the work is the size and variation of the data set we used. All the patterns were created under similar laboratory conditions. Although some experimental settings like the velocity of the blood droplets were varied, under each setting only a few patterns were generated. Therefore, we believe the data is not very representative of crime scenes in the real world.

Another limitation is that we used a relatively simple model to focus on the application of likelihood ratio framework. Using features to represent bloodstain patterns inevitably yields a loss of information. Incorporating more features to help separate patterns of different mechanisms is one approach to improvement. But this requires clever design of characteristic features and it becomes more difficult as the number of possible mechanisms goes up. In addition, the assumption of a Gaussian distribution was driven primarily by convenience for this conceptual demonstration. A mixture model or kernel-based model \citep{lindsay1995mixture,jones1996brief} can be used to more flexibly describe complex data distributions. Of course, a more flexible model with high-dimensional input space can easily lead to overfitting and produce extreme values of the likelihood ratio. Even with our relatively simple model, the absolute value of the logarithm of the likelihood ratio can be large. Possible solutions to reduce overfitting include parameter regularization, hierarchical modeling, and collecting more data.

Despite the limitations, the results show the potential for application of the likelihood ratio approach in bloodstain pattern analysis. In future study we will focus on collecting more representative data and building more flexible statistical models to better estimate the likelihood of bloodstain patterns.

\appendix
\section{Appendix}
\label{Appendix}
For cases where there are more than two possible hypotheses regarding the bloodstain pattern evidence $E$, formula (\ref{lrdef}) can be generalized to evaluate the strength of the evidence supporting one hypothesis against the others. Assume there are $K$ mutually exclusive hypotheses $H_1$, $H_2$,$...$,$H_{K}$ proposing different possible mechanisms for $E$. Without loss of generality, let $H_1$ be the main hypothesis to be considered against the general alternative ``some other mechanisms'', then the LR can be written as:
\begin{align}
LR = \frac{P(E|H_1)}{P(E|H_2\text{ or }M_3\text{ or ... or }H_K)} = \frac{P(E|H_1)}{P(E|\cup_{i=2}^K H_i)} \label{lrm}
\end{align}
According to the definition of conditional probability, the denominator can be organized as:
\begin{align}
P(E|\cup_{i=2}^K H_i) &= \frac{P(E\cap(\cup_{i=2}^K H_i))}{P(\cup_{i=2}^K H_i)} \label{lr1} \\
&= \frac{\sum_{i=2}^K P(E\cap H_i)}{\sum_{i=2}^K P(H_i)} \\
&= \frac{\sum_{i=2}^K P(E|H_i)P(H_i)}{\sum_{i=2}^K P(H_i)} \\
&= \sum_{i=2}^K w_i P(E|H_i) \label{lr2}
\end{align}
where $w_i=P(H_i)/\sum_{i=2}^K P(H_i)$ are prior probability weights for the other hypotheses. This result shows that evaluating $P(E|H_i)$, the likelihood of a pattern under each hypotheses $H_i$, remains the key calculation that is required.

\printcredits

\bibliographystyle{cas-model2-names}

\bibliography{cas-refs}


\end{document}